\documentstyle[amssymb,aps]{revtex}

\begin{document}
\title{On a biphononic origin of the 1125 cm$^{-1}$ absorption band in cuprous oxide%
\bigskip\ }
\author{V.M.Burlakov$^{2\text{ \#}}$, M.G\"{o}ppert$^1$, A.Jolk$^1$, A.Dinger$^1$,
R.Becker$^1$, and C.F.Klingshirn$^1$\bigskip}
\address{$^1$Institut f\"{u}r Angewandte Physik, Universit\"{a}t Karlsruhe,\\
Kaiserstr.12, D - 76128 Karlsruhe, Germany\\
$^2$Institute for Spectroscopy Russian Academy of sciences, 142092 Troitsk,\\
Russia \bigskip\ \bigskip\ }
\maketitle

\begin{abstract}
We report on\ the IR spectroscopic studies in both reflection ($50-900$ cm$%
^{-1}$) and transmission ($900-3000$ cm$^{-1}$) mode of the vibration
spectrum of the cuprous oxide. A detailed analysis based on a comparison of
the temperature dependences of the absorption band at $\simeq $1125 cm$^{-1}$
and of IR and Raman active fundamental vibrations results in assignment of
the former to a biphonon.

\bigskip\ PACS numbers: 61.82.Fk, 78.30.-j, 63.20.Ry

\bigskip\ $^{\#}$The author to whom correspondence should be addressed.
E-mail: burlakov@isan.troitsk.ru\ 
\end{abstract}

\section{Introduction}

Cuprous oxide ($Cu_2O$) has been studied for many years, mainly because of
its very rich electronic spectrum showing a large number of excitonic
structures associated with quadrupole and dipole transitions and
phonon-assisted dipole transitions \cite{review,Froelich,Jolk}. It serves
also as a model object for investigation of nonequilibrium phenomena in
electronic and excitonic systems \cite{cardona}. Considerable attention has
been paid to the phonon spectrum as well. Despite numerous infrared and
Raman studies the phonon spectrum of cuprous oxide is still not well
understood. The main problem is that the observed number of bands is much
larger than that deduced from the group theory. This fact lead the authors
to interpret the spectra in terms of models involving nonstochiometry or
impurity defects or multiphonon processes in light absorption (scattering).
The most intriguing point in the IR spectrum concerns the absorption peaks
at frequencies higher than 1000 cm$^{-1}$. The band at $\sim $1125 cm$^{-1}$%
, for instance, was assigned to the transitions between two different sets
of exciton levels \cite{ref1a}, i.e. between the so-called ''yellow'' and
''green'' exciton series \cite{ref1b,ref1c,ref1d}. It was supposed that
during the IR measurements the sample is illuminated with white light
resulting in a population of long-living excitonic states which could absorb
the infrared quanta. In Ref. \cite{ref2} the 1125 cm$^{-1}$ feature was
assigned to a fundamental lattice vibration, though no indications of such a
high frequency fundamental vibration are available from inelastic neutron
scattering studies \cite{netron} and lattice dynamics calculations \cite
{latdyn}. Another possible explanation of the 1125 cm$^{-1}$ band was given
in Ref. \cite{ref1e}, where it was assigned to silicon-like impurities. In
Ref. \cite{ref1} it was supposed that this band is due to multiphonon
process. Thus, no convincing interpretation of the high frequency peak at $%
\sim $1125 cm$^{-1}$ has been given so far.

Present study is aimed at the origin of the $\sim $1125 cm$^{-1}$ feature in
the absorption spectrum of $Cu_2O$. To exclude its electronic origin we
examined an influence of laser illumination on the absorption band intensity
at $\sim $1125 cm$^{-1}$. Comparative study of the temperature dependences
of the peak frequency and the band halfwidth on one hand and those of the IR
and Raman active fundamental vibrations on the other hand showed that the
absorption band at $\sim $1125 cm$^{-1}$ is of a one-particle nature. Its
peak position possesses quite different temperature dependence than any
simple combination of the fundamental vibration frequencies. Also the
temperature dependence of the band halfwidth is dominated by participation
of high frequency ($\omega \gtrsim 500\quad cm^{-1}$) phonons in the
corresponding anharmonic decay processes. Thus, the $\sim $1125 cm$^{-1}$
feature can be assigned either to some sort of intrinsic defect or to a
biphonon, i.e. bound state of two phonons splitted from the two-phonon band $%
\omega _0(k)+\omega _0(-k)$ \cite{biphonon}. Biphonon is a single particle
excitation and, disregarding rather unlike defect origin of the 1125 cm$%
^{-1} $ feature, it can account for the peculiarities of the high frequency
absorption band.

\section{Experimental}

All the spectral measurements to be described were carried out with a Bruker
113v Fourier-transform spectrometer in the range 50$-$3000 cm$^{-1}$ with a
spectral resolution of 0.5 cm$^{-1}$. Reflectance measurements were made at
an angle of incidence 11$^{\circ }$ on a bulk single crystal of 3x3 mm$^2$
area while the transmission was measured on a 1.5x1.5 mm$^2$ single
crystalline sample of 70 $\mu $ thickness. The samples were cut and polished
from high quality naturally grown single crystal.

For the low temperature measurements the sample was mounted in a special
holder in a helium flow cryostat with temperature control better than 0.5 K.
The reflectivity data were transformed into a dielectric function via
Kramers-Kronig transformation and the peak maxima and halfwidths of the
spectra of $Im[\varepsilon (\omega )]$ and $Im[-\varepsilon ^{-1}(\omega )]$
were analyzed.

\section{Results and discussion}

The general features of the IR reflectance spectrum of $Cu_2O$ have been
published by several authors \cite{ref4,teylor}. It is well established that
there are two allowed IR active phonons of $F_{1u}$ symmetry with TO phonon
frequencies $\omega _{TO1}\simeq 150$ cm$^{-1}$ and $\omega _{TO2}\simeq 605$
cm$^{-1}$. The assignments of the absorption peaks have been also given \cite
{ref3,ref7}. All the assignments made so far were based on the symmetry
considerations when concerning the fundamental modes (the isotope shift of
oxygen was also used in the assignment of the $\simeq 605$ cm$^{-1}$ IR band 
\cite{ref2}) and also simple combinations of the fundamental mode
frequencies were applied to multiphonon peaks. The symmetry arguments for
the latter ones are, however, not so strong as for the fundamental
vibrations. A weaker intensity of multiphonon bands can be caused either by
a small dipole moment for dipole allowed combinations of phonons or by a
large electric quadrupole moment for dipole forbidden combinations.
Additionally, the frequency of the multiphonon absorption band is not
necessary exactly equal to a combination of phonon frequencies at the $%
\Gamma $ point of the Brillouin zone. Thus, the assignment of a relatively
weak IR band might be rather ambiguous. To make it more convincing one can
consider the temperature dependent spectra assuming that the multiphonon
peaks are combined from one and the same phonons over a wide temperature
range. Then, obviously, the multiphonon peak position must follow the
combination of those of the constituent phonons when heating or cooling the
sample. The integrated intensity of the multiphonon peaks is generally also
temperature dependent, but its determination is usually not of sufficiently
high accuracy and can not be used as an argument in favor or against their
multiphonon nature.

\subsection{The 1125 cm$^{-1}$ feature is not of electronic origin}

Fig.1 presents the spectra of optical density obtained from the
transmittance measured at various temperatures. The well pronounced and well
separated absorption band at $\omega _B\simeq $1125 cm$^{-1}$ (at T=295 K)
is of our specific interest. The band shows a clear red shift and broadening
with increasing temperature. Its peak position and halfwidth are plotted in
Figs.2 and 3 respectively. There are also a number of overlapping bands at
lower frequencies the parameters of which are not presented because of the
low accuracy in their determination. By open circles we show in Fig.1 also
the optical density of the sample illuminated by an Ar$^{+}$ cw laser ($%
\lambda =514$ nm) with power about 1 W focused on 2 mm$^2$ at a nominal
sample temperature of 10 K. Obviously the electronic excitation efficiency
for the illumination conditions used is many orders in magnitude higher than
that for illumination during the ordinary IR measurements and should result
in a remarkable change of the $\omega _B$ band intensity if it is of the
electronic origin discussed in \cite{ref1a}. Actually one sees from Fig.1
that the illumination results only in a small red shift and a broadening of
the band just by the sample heating effect. No indications of absorption
related to transitions between excitonic level is observed. The small peaks
at $\simeq $1000 cm$^{-1}$ and $\simeq $1250 cm$^{-1}$ (curve 4 in Fig.1)
remain unchanged after switching off the illumination suggesting some sort
of persistent photoinduced origin. Together with the isotope shift of the
frequency $\omega _B$ \cite{ref2} our data give strong evidence for a {\it %
pure vibrational nature }of the band under discussion.

\subsection{The 1125 cm$^{-1}$ feature is not a multiphonon band}

To answer the question about the multiphonon nature of the $\omega _B$ band
we measured the temperature dependent reflectance spectrum (see Fig.4). The
shifts of the TO and LO phonon frequencies obtained via Kramers-Kronig
transformation from the reflectance spectra are shown in Fig.5. The
frequency shift of the Raman line at $\omega _R\simeq $515 cm$^{-1}$\cite
{ivanda} which in combination with $\omega _{TO2}$ could result in an
absorption band at $\sim $1125 cm$^{-1}$ is also presented. The experimental
temperature dependences of the frequency shifts were fitted with the
function 
\begin{equation}
\Delta \omega (T)=\omega (T=10K)-\omega (T)=\Delta _0+C_1\cdot T+C_2\cdot
T^2,  \label{eq1}
\end{equation}
corresponding to the temperature dependences of the phonon band frequencies
due to phonon decay processes and due to thermal expansion of the crystal
lattice. $\Delta _0,$ $C_1$ and $C_2$ in (\ref{eq1}) are fitting parameters.
It turned out that the frequency shift of the $\omega _B$ band can be
represented neither as the combination $\Delta \omega _B=\Delta \cdot
(3\cdot \omega _{LO1}+\omega _{LO2})$ \cite{ref1} giving rise for electric
quadrupole absorption nor as the $\Delta \omega _B=\Delta \cdot (\omega
_R+\omega _{TO2})$ suggesting electric dipole absorption \cite{ziomek}.
Moreover, no reasonable fit can be reached at all using Eq. (\ref{eq1}) (see
dotted line in Fig.2). Under a quite realistic assumption about the validity
of the Eq. (\ref{eq1}) over the whole Brillouin zone this fact means that
the $\omega _B$ band is {\it not a simple multiphonon} one.

\subsection{What is the origin of the $\omega _B$ band?}

Since the $\omega _B$ band corresponds neither to an electronic transition
nor to a multiphonon process one may turn back to its defect or impurity
origin. It seems however quite unlikely that the relative intensity of the
defect or impurity band compared to the fundamental absorption bands is
nearly the same in a variety of samples from different sources and after
different treatment. Therefore one should consider some sort of intrinsic
defect, such as an oxygen atom (as evidenced by the isotope effect \cite
{ref2}) in an interstitial site (see also the argumentation in \cite{ref3}).
Even in this case, however, it is difficult to explain such a high frequency
of an oxygen vibration which normally does not exceed 800 cm$^{-1}.$

The temperature dependences of the frequency shift $\Delta \omega _B(T)$ and
the halfwidth $S_B(T)$ of the $\omega _B$ band can be described nearly
perfectly by anharmonicity of third and fourth order \cite{reiss} involving
the high frequency phonons only suggesting a one-phonon origin of the $%
\omega _B$ band. Figs 2 and 3 (solid lines) demonstrate the fit obtained
according to the expressions 
\begin{equation}
\begin{array}{c}
\Delta \omega _B(T)=\omega _B(T=0)-\omega _B(T)\sim C_1\cdot T+C_2\cdot
(n(\omega _{ph},T)+ \\ 
n(\omega _B(T=0)-\omega _{ph},T))+C_3\cdot n(\omega _{ph},T)\cdot \left(
n(\omega _{ph},T)+1\right) ,
\end{array}
\label{eq2a}
\end{equation}

\begin{equation}
\begin{array}{c}
S_B(T)\sim S_B(T=0)\cdot (1+n(\omega _{ph},T)+n(\omega _B(T=0)-\omega
_{ph},T))+ \\ 
C_4\cdot n(\omega _{ph},T)\cdot \left( n(\omega _{ph},T)+1\right) ,
\end{array}
\label{eq2b}
\end{equation}
where $n(\omega ,T)=(\exp (\hbar \omega /k_BT)-1)^{-1}$ is phonon occupation
number, $\omega _B(T=0)=1153$ cm$^{-1}$, $S_B(T=0)\simeq 68$ cm$^{-1}$, $%
\omega _{ph}=\omega _B/2.$ Constants $C_1,$ $C_2,$ $C_3,$ and $C_4$ are
fitting parameters. The thermal expansion effect is taken into account
through the constant $C_1$. Dotted lines in Figs 2 and 3 show the best fits
using Eq (\ref{eq1}) for the $\Delta \omega _B(T)$ and Eq (\ref{eq2b}) with $%
\omega _{ph}\lesssim \omega _B/4$ for the $S_B(T)$, e.g. assuming that the $%
\omega _B$ band is of a two-phonon origin. Note, that the total number of
the fit parameters in the latter case is even higher than that in the former
one.

From the captures to Figs 2 and 3 one can see that the fourth order
anharmonicity gives considerable contribution to both broadening and
frequency shift of the $1153$ cm$^{-1}$ vibration. Now two important facts
have to be pointed out: i) the value of the phonon frequency $\omega
_{ph}\simeq 600$ cm$^{-1}$ appears to be very close to $\omega _{TO2}$; ii)
the isotope effects for the $\omega _B$ and $\omega _{TO2}$ vibrations are
nearly the same \cite{ref2}. Both two facts can be naturally incorporated in
the picture which considers the $\omega _B$ vibration as a bound state of
two $\omega _{TO2}$ phonons, i.e. a biphonon. Indeed, fact i) means a much
stronger interaction of the $\omega _B$ vibration with the $\omega _{TO2}$
phonons than with any other phonons just as a consequence of strong
interaction between the $\omega _{TO2}$ phonons themselves resulting in the
biphonon formation. The fact ii) gives evidence for a similarity between the
eigenvectors of the $\omega _B$ and of the $\omega _{TO2}$ vibrations
suggesting their common origin. Within the biphonon picture the IR activity
of $\omega _B$ must be due to an electric quadrupole transition.

The biphonon picture described implies rather low binding energy $%
E_{bind}\simeq \hbar \cdot (2\omega _{TO2}-\omega _B)\simeq $100 cm$^{-1}$
of the biphonon. This will result in a dissociation of the biphonons, i.e.
in an additional broadening of the $\omega _B$ band at elevated temperatures
caused by increasing interaction with phonons $\omega \simeq E_{bind}$, e.g.
acoustic phonons. The process responsible for such a broadening would give
the contribution

\begin{equation}
\Delta S_B(T)\sim \left[ n(E_{bind},T)\cdot \left( n(\omega
_{TO2},T)+1\right) ^2-n^2(\omega _{TO2},T)\cdot \left(
n(E_{bind},T)+1\right) \right] .  \label{eq3}
\end{equation}
As the $\omega _B$ band broadening does not contain any noticeable
contribution of the low frequency phonons one may conclude that neither the
process (\ref{eq3}) nor those considered in \cite{wess} are not essential in
our case.

\section{Conclusion}

We performed IR spectroscopic studies both in reflection and transmission
modes of the vibration spectrum of cuprous oxide. Detailed analysis of the
temperature behavior of the resonance frequency and of the halfwidth of the
absorption band at $\omega _B\simeq $1125 cm$^{-1}$ on one hand and of the
frequencies of IR and Raman active phonons on the other hand result in the
conclusion about a biphononic origin of the 1125 cm$^{-1}$ vibration.

\section{Acknowledgments}

The authors are grateful to Prof. B. Mavrin for valuable comments. We would
like to thank the Deutsche Forschungsgemeinschaft for financial support of
our work. One of the authors (V.M.B.) gratefully acknowledges support from
the Graduiertenkolleg ''Kollektive Ph\"{a}nomene im Festk\"{o}rper'' during
his recent visits at the Universit\"{a}t Karlsruhe.

\begin{center}
\newpage Figure captures
\end{center}

\begin{enumerate}
\begin{description}
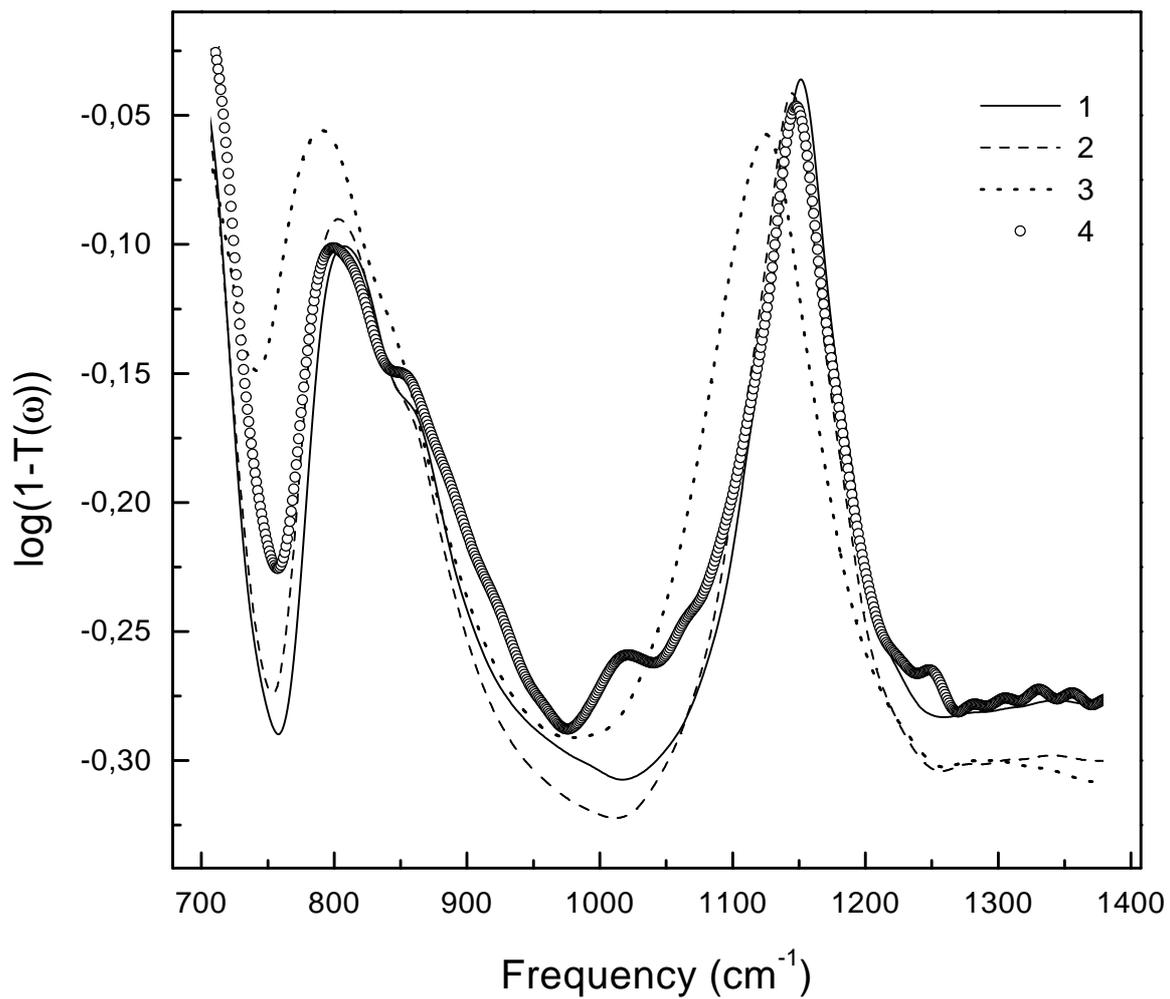
\item  Fig.1. Spectra od the optical density of cuprous oxide at temperature
10 K (1), 200 K (2), 295 K (3), and of 10 K with laser illumination (see
text) (4).

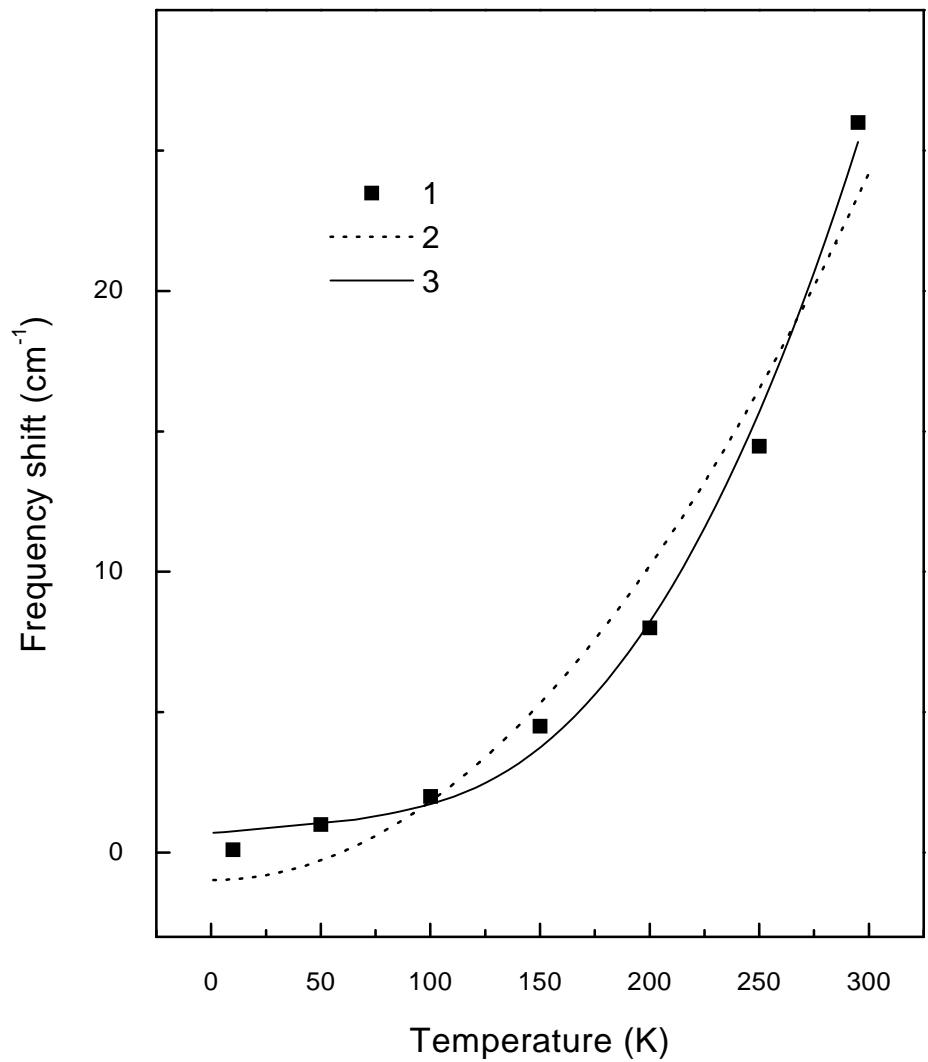
\item  Fig.2. Frequency shift $\Delta \omega $ of the $\omega _B\simeq 1125$
cm$^{-1}$ absorption band in cuprous oxide versus temperature. (1)
experimental data; (2) best fit with Eq. (\ref{eq1}); (3) fit with Eq. (\ref
{eq2a}) with $C_1=0.013$ cm$^{-1}/K,$ $C_2=104.16$ cm$^{-1},$ $C_3=103.81$ cm%
$^{-1}$.

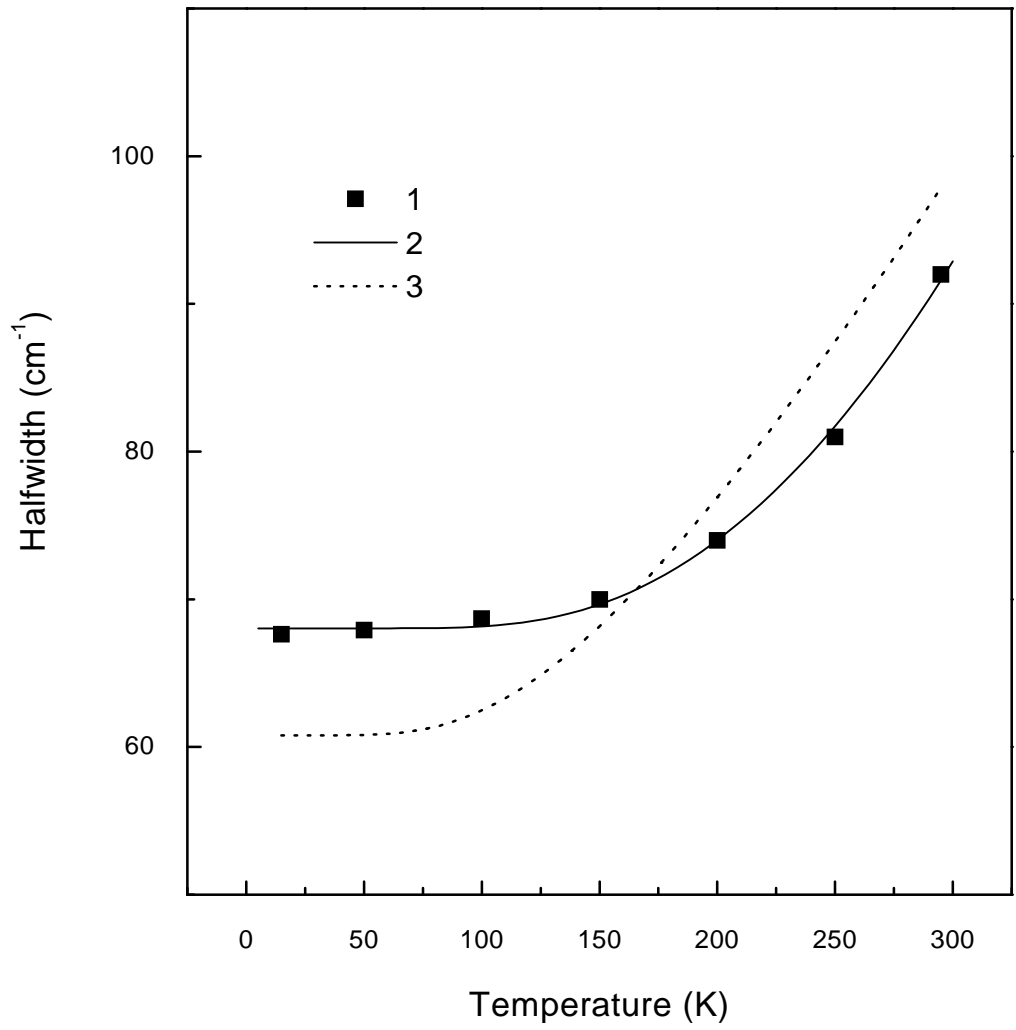
\item  Fig.3. Halfwidth of the $\omega _B\simeq 1125$ $cm^{-1}$ absorption
band in cuprous oxide versus temperature. (1) experimental data; (2) fit
with Eq (\ref{eq2b}) with $C_4=208.35$ cm$^{-1}$; (3) see text.

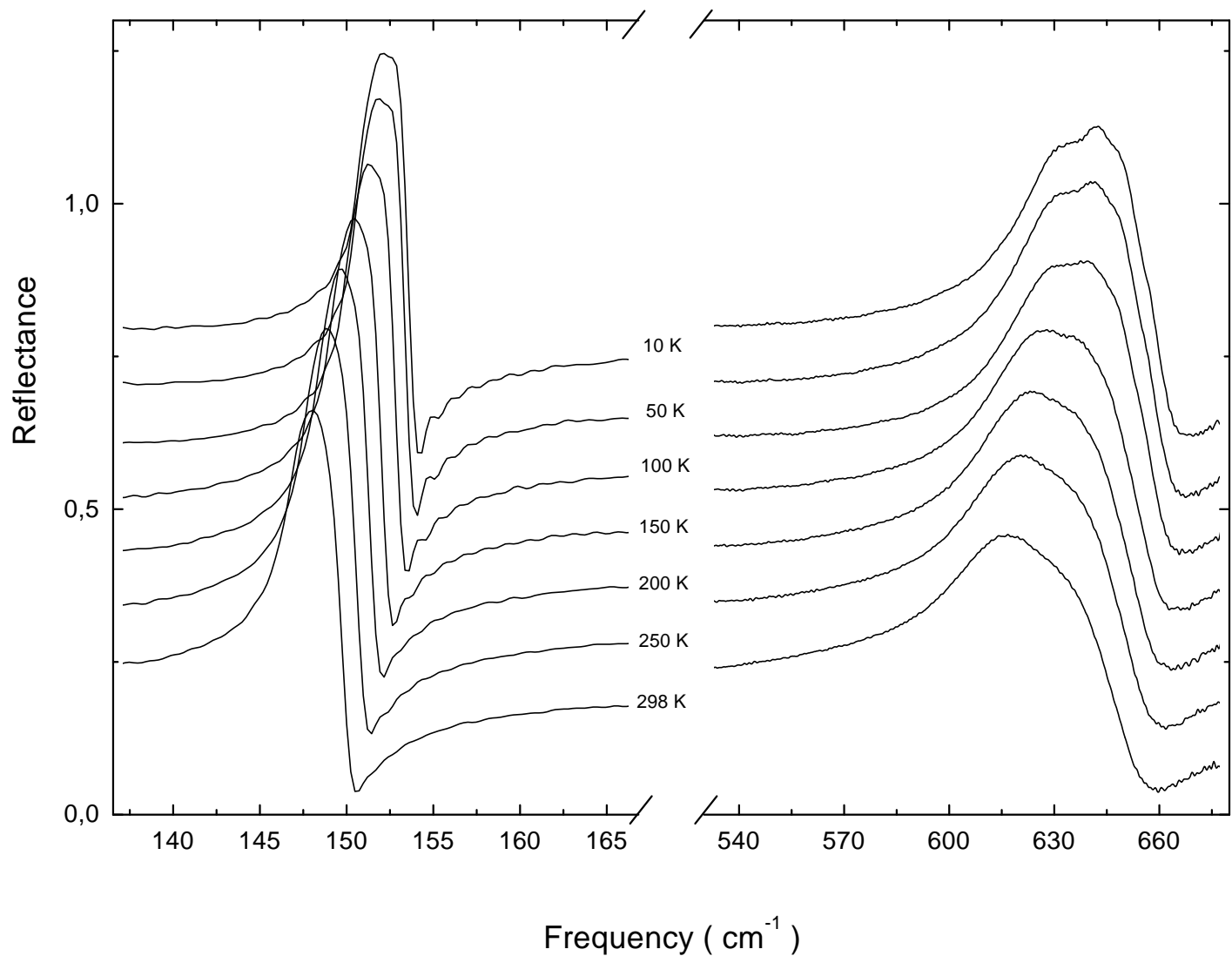
\item  Fig.4. Reflectance spectra of cuprous oxide for various temperatures.

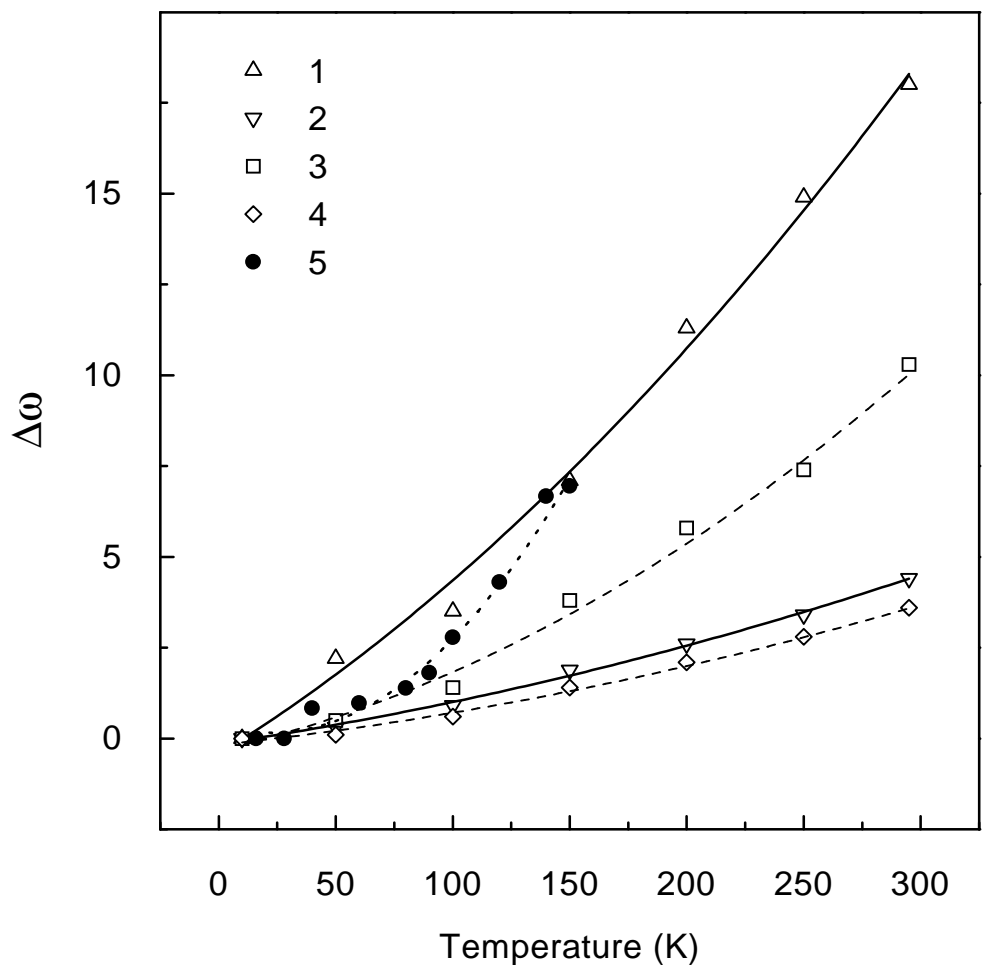
\item  Fig.5. Frequency shifts $\Delta \omega $ of the fundamental
vibrations in cuprous oxide versus temperature. (1) $\omega _{TO2}\simeq 605$
cm$^{-1};$ (2) $\omega _{TO1}\simeq 150$ cm$^{-1}$; (3) $\omega _{LO2}\simeq
660$ cm$^{-1}$; (4) $\omega _{LO2}\simeq 153$ cm$^{-1}$; (5) Raman line at $%
\omega _R\simeq 515$ cm$^{-1}$ \cite{ivanda}.
\end{description}
\end{enumerate}

\end{document}